\documentclass[twocolumn,english,aps,prl,superscriptaddress]{revtex4-1}
\usepackage[T1]{fontenc}
\usepackage[latin9]{inputenc}
\usepackage{amsmath}
\usepackage{amssymb}
\usepackage{graphicx}
\usepackage{esint}

\makeatletter
\@ifundefined{textcolor}{}
{%
 \definecolor{BLACK}{gray}{0}
 \definecolor{WHITE}{gray}{1}
 \definecolor{RED}{rgb}{1,0,0}
 \definecolor{GREEN}{rgb}{0,1,0}
 \definecolor{BLUE}{rgb}{0,0,1}
 \definecolor{CYAN}{cmyk}{1,0,0,0}
 \definecolor{MAGENTA}{cmyk}{0,1,0,0}
 \definecolor{YELLOW}{cmyk}{0,0,1,0}
 }

\usepackage{babel}

\usepackage{tikz}
\newcommand*{\circled}[1]{\tikz[baseline=(char.base)]{
            \node[shape=circle,draw,inner sep=1pt] (char) {#1};}}

\usepackage{babel}

\makeatother

\usepackage{babel}
\begin{document}

\preprint{This line only printed with preprint option}

\title{Taming the dynamical sign problem in real-time evolution of quantum
many-body problems}

\author{Guy Cohen}

\affiliation{Department of Chemistry, Columbia University, New York, New York
10027, U.S.A.}

\affiliation{Department of Physics, Columbia University, New York, New York 10027,
U.S.A.}

\author{Emanuel Gull}

\affiliation{Department of Physics, University of Michigan, Ann Arbor, MI 48109,
U.S.A.}

\author{David R. Reichman}

\affiliation{Department of Chemistry, Columbia University, New York, New York
10027, U.S.A.}

\author{Andrew J. Millis}

\affiliation{Department of Physics, Columbia University, New York, New York 10027,
U.S.A.}
\begin{abstract}
Current nonequilibrium Monte Carlo methods suffer from a dynamical
sign problem that makes simulating real-time dynamics for long times
exponentially hard. We propose a new `Inchworm Algorithm', based on
iteratively reusing information obtained in previous steps to extend
the propagation to longer times. The algorithm largely overcomes the
dynamical sign problem, changing the scaling from exponential to quadratic.
We use the method to solve the Anderson impurity model in the Kondo
and mixed valence regimes, obtaining results both for quenches and
for spin dynamics in the presence of an oscillatory magnetic field. 
\end{abstract}
\maketitle
The nonequilibrium physics of quantum many-body systems is a central
topic of current research \cite{eisert_quantum_2015}. Experimentally,
the application of strong currents through quantum dots \cite{van_der_wiel_electron_2002},
molecular junctions \cite{zimbovskaya_electron_2011} and extended
systems, the optical excitation of high densities of carriers above
band gaps of Mott insulators \cite{dienst_optical_2013} and high
amplitude terahertz coupling to phonon modes \cite{liu_terahertz-field-induced_2012}
have revealed exciting new physics. In the cold atom context `quenches'
(sudden changes of parameters) have also been extensively studied
\cite{sadler_spontaneous_2006,chen_quantum_2011,gring_relaxation_2012}.
While remarkable experimental progress has been made, theory faces
a crucial limitation: numerical calculations of time-dependent and
nonequilibrium problems suffer from an exponential scaling of computational
cost with simulation time. In different formulations of nonequilibrium
calculations the problem manifests itself in different ways: for instance,
as a mixing of low- and high-energy states as time progresses in truncated
wavefunction methods like time dependent NRG \cite{anders_real-time_2005}
or DMRG \cite{white_real-time_2004}, as an exponential number of
operators needed to reach a given accuracy in the hierarchical equations
of motion \cite{tanimura_stochastic_2006,welack_influence_2006,jin_exact_2008,hartle_decoherence_2013,hartle_transport_2015},
or as a `dynamical' sign problem in nonequilibrium quantum Monte Carlo
(QMC) \cite{muhlbacher_real-time_2008,werner_diagrammatic_2009,schiro_real-time_2010,antipov_voltage_2015}.
In practice, the exponential scaling of the known numerically exact
methods has prevented accurate numerical calculations of the long-time
behavior of nonequilibrium correlated systems.

Diagrammatic QMC methods, which provide numerically exact solutions
by stochastically sampling a perturbation series, have been particularly
fruitful in elucidating the physics in equilibrium, where the problem
can be formulated in imaginary time so that the calculation concerns
the estimation of combinations of decaying exponentials \cite{prokofev_worm_1998,rubtsov_continuous-time_2005,werner_continuous-time_2006,prokofev_bold_2007,prokofev_bold_2008,gull_continuous-time_2008,gull_continuous-time_2011,semon_importance_2012,shinaoka_negative_2015}.
The straightforward extension of diagrammatic QMC methods to the nonequilibrium
situation \cite{muhlbacher_real-time_2008,werner_diagrammatic_2009,schiro_real-time_2010,antipov_voltage_2015}
requires estimation of integrals that contain combinations of oscillating
exponentials $\exp(iHt)$; as the integrals extend over longer time
ranges, numerical difficulties limit the times accessible in the strong
coupling regime to the order of the typical tunneling timescale. Longer
times can be reached by sampling corrections to semi-analytic theories
such as the non-crossing approximation (NCA) \cite{gull_bold-line_2010,gull_numerically_2011}
and the one-crossing approximation (OCA) \cite{cohen_numerically_2013,cohen_greens_2014,cohen_greens_2014-1},
by explicit summation over Keldysh indices followed by a continuation
on the complex plane \cite{profumo_quantum_2015}, and by using memory
function techniques to continue very precise short-term results \cite{cohen_memory_2011,cohen_generalized_2013,wilner_bistability_2013,wilner_nonequilibrium_2014}.
Nevertheless, the basic problem of dealing with oscillating exponentials
remains, so that all of these methods encounter an exponential wall
as time is increased, limiting their applicability to relatively short
time dynamics or to the weak correlation regime.

In this Letter we present a solution to this problem in terms of a
new algorithm whose computational cost scales \emph{quadratically}
rather than exponentially with time, allowing controlled numerical
access to the long-time behavior of strongly correlated quantum systems.
The algorithm is based on iteratively reusing information from shorter
time propagation to obtain results for longer times, is generally
applicable to any diagrammatic method and has a straightforward interpretation
in terms of self-consistent skeleton expansions. The method presented
here deals only with the dynamical sign problem, not with the intrinsic
fermionic one, which limits access to certain systems even in equilibrium.
However, it is possible to also conceive of a spatial inchworm algorithm
(as opposed to the temporal one presented here) which might make headway
against that problem. We present an implementation of the algorithm
for a paradigmatic quantum many-body system, the Anderson impurity
model, in the strongly correlated Kondo and mixed-valence regimes,
and show that it captures the long-time spin dynamics after a quantum
quench and in the presence of an oscillating magnetic field. While
the results presented here pertain to impurity models, the algorithm
itself should prove useful beyond this context in the more general
quantum many-body setting.

\begin{figure}
\includegraphics[width=8.6cm]{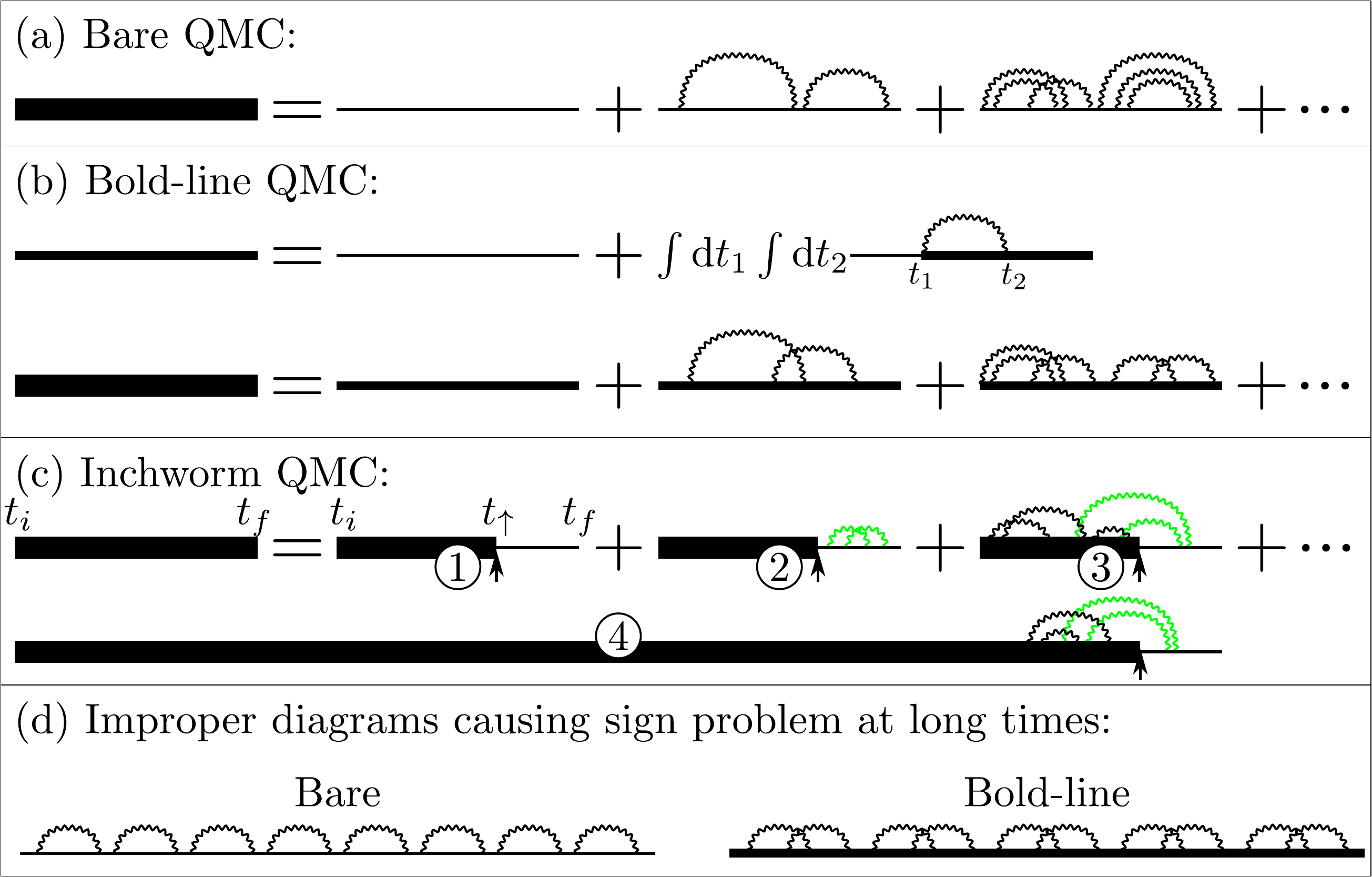}\caption{Comparison of diagrams sampled in previous approaches (bare expansion,
panel (a) \cite{muhlbacher_real-time_2008,werner_diagrammatic_2009}
and bold expansion, panel (b) \cite{gull_bold-line_2010,gull_numerically_2011})
to diagrams sampled in new approach (c), along with examples of diagrams
leading to dynamical sign problem in previous methods (d). Thick lines:
full propagators. Thin lines: bare propagators. Medium (or `bold')
lines: propagators resulting from analytical resummation of subset
of diagrams (here, NCA). Wiggly lines: hybridization lines. Arrows
indicate $t_{\uparrow}$.\label{fig:method_diagrams}}
\end{figure}

The crucial object in the algorithm is the Keldysh-contour propagator
$G_{\alpha\alpha^{\prime}}(t_{1},t_{2})$ giving the transition amplitude
between state $\alpha$ at contour time $t_{i}$ and state $\alpha^{\prime}$
at contour time $t_{f}$ in the presence of a Hamiltonian $H=H_{0}(t)+V(t)$:
\begin{equation}
G_{\alpha\alpha^{\prime}}\left(t_{f},t_{i}\right)\equiv\left\langle \alpha\right|\mathrm{Tr}_{\mathcal{B}}\left\{ e^{i\int_{t_{i}}^{t_{f}}\mathrm{d}\tilde{t}H_{0}(\tilde{t})+V(\tilde{t})}\right\} \left|\alpha^{\prime}\right\rangle .\label{eq:propagator_definition}
\end{equation}

Here $H_{0}$ is assumed to be an exactly solvable Hamiltonian and
one studies $G$ by an expansion in $iV$, as illustrated for an impurity
model expansion (where all propagators can be collapsed onto a single
line) in the top two panels of Fig.~\ref{fig:method_diagrams}. Panel~(a)
represents a bare expansion, where $G$ (thick line) is evaluated
by summing all possible interaction lines in terms of a bare propagator
(thin line). Panel~(b) represents a particular bold-line expansion,
where an approximate propagator (represented by a medium or `bold'
line) containing a subset of the interactions is evaluated semi-analytically,
and all corrections to that approximation are summed of in terms of
the bold propagator. It is important to note that $G$ is contour
causal: in the expansion only vertices $V(\tilde{t})$ for which $t_{i}<\tilde{t}<t_{f}$
occur. The factors of $iV$ cause a dynamical sign problem, and in
the approaches used to date the expansion order (number of insertions
of $iV$) is proportional to the final time simulated. Our new algorithm
avoids this problem by using a kind of skeleton diagram expansion:
it exploits the contour causal nature of G to construct an exact propagator
for longer times in terms of an exact propagator for shorter times,
iteratively increasing the time up to which propagators are known.
In practice, we observe that the sign problem does not worsen as a
function of time, resulting in an overall quadratic algorithmic scaling.

The algorithm, which we illustrate in Fig.~\ref{fig:method_diagrams}
(c), begins from the assumption that $G_{\alpha\alpha^{\prime}}\left(t_{1,}t_{2}\right)$
is known for all values of $t_{1}$ and $t_{2}$ less than a designated
time $t_{\uparrow}$. We now consider the terms appearing in a computation
of $G_{\alpha\alpha^{\prime}}\left(t_{f},t_{i}\right)$ for $t_{f}>t_{\uparrow}$.
If no interactions occur or all interactions occur before $t_{\uparrow}$,
the term can be subsumed into the (known) propagation from $t_{i}$
to $t_{\uparrow}$, followed by a bare propagation from $t_{\uparrow}$
to $t_{f}$, as illustrated in diagram \circled{1} of Fig.~\ref{fig:method_diagrams}~(c).
If interactions occur after $t_{\uparrow}$ but no interaction lines
connect times after $t_{\uparrow}$ to times before $t_{\uparrow}$,
the propagation to $t_{\uparrow}$ is captured by the known $G_{\alpha\alpha^{\prime}}\left(t_{\uparrow},t_{i}\right)$,
with the usual perturbation in $V$ required to capture propagation
in the interval $t_{\uparrow}\rightarrow t_{f}$ (see diagram \circled{2}).
Finally, terms with interaction lines spanning $t_{\uparrow}$ can
be subsumed into diagrams with exact propagators before $t_{\uparrow}$
and bare propagators after $t_{\uparrow}$ by absorbing any interaction
line that is not connected to a line reaching past $t_{\uparrow}$
in the exact propagator (diagram \circled{3}).

By summing these three classes of diagrams (\circled{1}, \circled{2},
\circled{3}) we count all possible diagrams exactly once, producing
a formally exact solution for the propagator $G_{\alpha\alpha^{\prime}}\left(t_{1,}t_{2}\right)$.
The procedure crucially relies on the contour-time causality of the
propagator: $G_{\alpha\alpha^{\prime}}\left(t_{1},t_{2}\right)$ contains
all possible diagrams with interaction lines between $t_{2}$ and
$t_{1}$ but no interaction lines outside of this interval.

The main difference with previously considered expansions is that
improper repetitions of simple inclusions (see panel (d) of Fig.~\ref{fig:method_diagrams})
are absorbed in the propagator for $t<t_{\uparrow}$ and only need
to be sampled for $t>t_{\uparrow}$. The number of these diagrams
grows exponentially as a function of propagation time, causing the
dynamical sign problem. To see this, one need only consider that the
number of possible locations for inclusions increases roughly linearly
with the length of the propagation time. Since each individual inclusion
might be removed, this generates an exponential number of possible
diagrams. $t_{\uparrow}$ is a free parameter: as $t_{\uparrow}$
is lowered to $t_{i}$, the procedure reverts to the standard bare
expansion in $V$ (see Fig.~\ref{fig:method_diagrams}(a)). As $t_{\uparrow}$
is increased towards $t_{f}$, fewer diagrams are sampled but the
exact propagator has to be known for longer times.

The possibility of obtaining propagators based on corrections to propagators
for smaller times suggests a numerical algorithm: starting from the
knowledge of the exact propagators within a short time interval $(t_{i},t_{f}^{n})$
with $t_{f}^{n}=t_{i}+n\Delta t$, e.g. as obtained from a bare Monte
Carlo simulation, we calculate propagators for the longer interval
$(t_{i},t_{f}^{n}+\Delta t)=(t_{i},t_{f}^{n+1})$ by setting $t_{\uparrow}=t_{f}^{n}$
and sampling again the three classes of diagrams described in Fig.~\ref{fig:method_diagrams}c.
The process is iteratively repeated, gradually increasing the interval
on which propagators are known by `inching' along the Keldysh contour.
These successive small steps which gradually increase $t_{f}$ have
led us to term this procedure the inchworm algorithm.

Since $G_{\alpha\alpha^{\prime}}\left(t_{f},t_{i}\right)$ has two
time arguments, propagation must be carried out in both temporal directions.
To reach a final time $t$ at a discretization of $\Delta t$ requires
$\frac{1}{4}\left(\frac{t}{\Delta t}\right)^{2}$ interdependent simulations
when causality and time-reversal symmetry are taken into account,
resulting in an algorithm that scales at least quadratically. To control
the complexity of the computation, it is also useful to limit the
maximum order of diagrams to be sampled and then verify convergence
with respect to increasing the diagram order \cite{prokofev_bold_2007,gull_numerically_2011}.
It can be shown that inchworm QMC truncated at a given order corresponds
as $\Delta t\rightarrow0$ to a self-consistent skeleton expansion
with the self-energy truncated to the same order. Based on experience
from these methods \cite{eckstein_nonequilibrium_2010} we may therefore
expect that most contributions at long times will include interaction
lines at only a limited, time-independent range from the final time,
as illustrated in diagram \circled{4} of panel (c) in Fig.~\ref{fig:method_diagrams}.

We illustrate the inchworm scheme with the example of an Anderson
impurity model with a time and spin dependent local field: 
\begin{eqnarray}
H\left(t\right) & = & \sum_{\sigma\in\left\{ \uparrow,\downarrow\right\} }\varepsilon_{\sigma}\left(t\right)d_{\sigma}^{\dagger}d_{\sigma}+Un_{\uparrow}n_{\downarrow}\label{eq:hamiltonian}\\
 &  & +\sum_{\sigma k}\varepsilon_{\sigma k}a_{\sigma k}^{\dagger}a_{\sigma k}+\sum_{a\sigma k}\left(V_{\sigma k}a_{\sigma k}^{\dagger}d_{\sigma}+\mathrm{H.C.}\right).\nonumber 
\end{eqnarray}
$\varepsilon_{\sigma}$ are on-site level energies, $\sigma\in\left\{ 1,-1\right\} $
a spin index, and $U$ the on-site Coulomb interaction. $\varepsilon_{\sigma k}$
and $V_{\sigma k}$ are fully defined by the dot--bath coupling, which
we set to a flat band with a soft cutoff: $\Gamma\left(\omega\right)$
$=$ $2\pi$ $\sum_{k}V_{\sigma k}^{*}V_{\sigma k}\delta\left(\omega-\varepsilon_{k}\right)$
$=$ $\Gamma$ $/$ $\left[\left(1+e^{\nu\left(\omega-\Omega_{c}\right)}\right)\left(1+e^{-\nu\left(\omega+\Omega_{c}\right)}\right)\right]$
with $\nu=10/\Gamma$ and $\Omega_{C}=10\Gamma$. $\Gamma$ will be
our unit of energy. We simulate a coupling quench, \emph{i.e.} the
dynamics of a dot initially decoupled from the bath, with the coupling
turned on instantaneously at time zero.

\begin{figure}
\includegraphics[width=8.6cm]{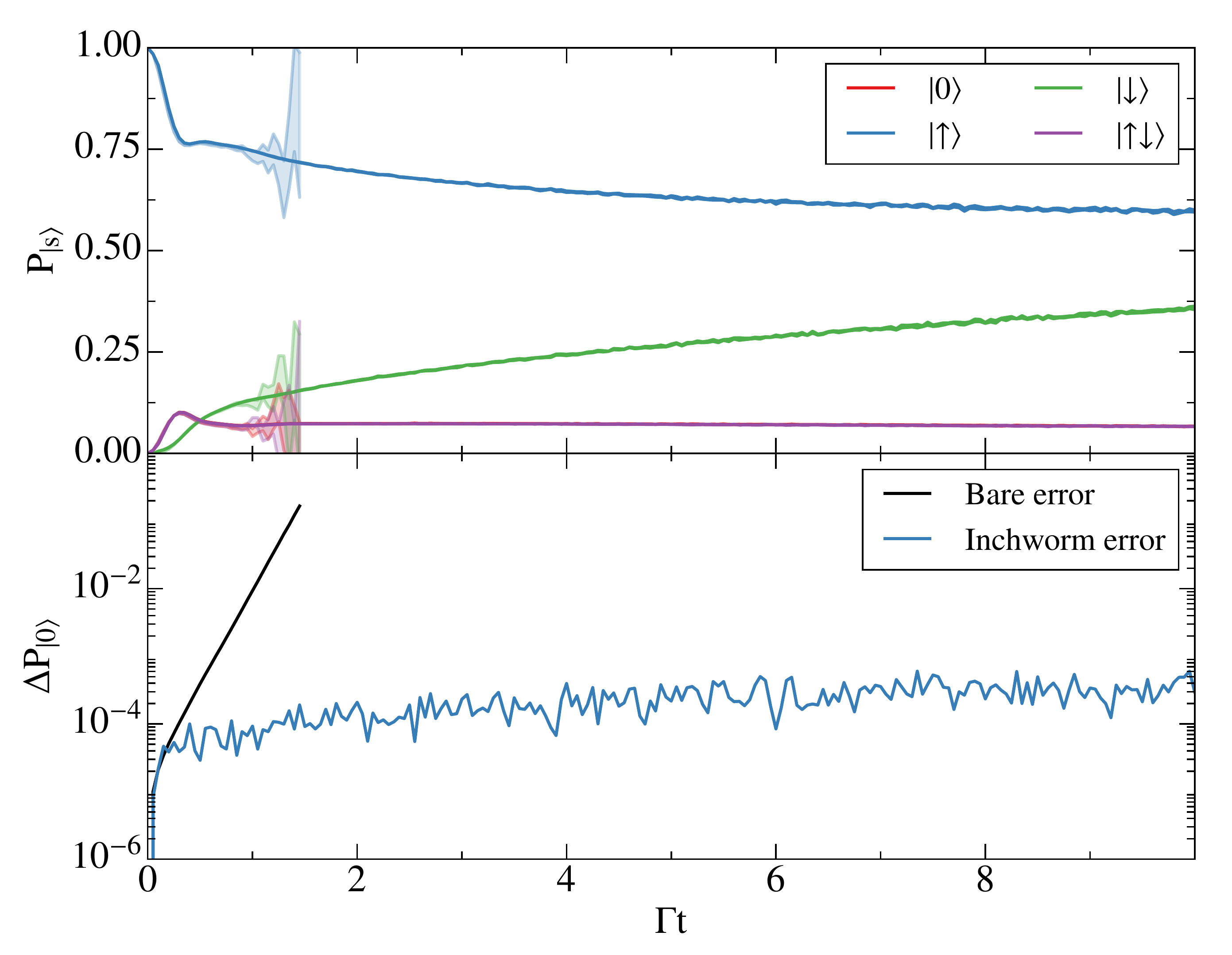}\caption{Top panel: population dynamics of the Anderson impurity model in the
Kondo regime following a coupling quench from a fully magnetized state
at $U=-2\varepsilon=8\Gamma$ and $\beta\Gamma=50$. The bare hybridization
expansion result (Fig. \ref{fig:method_diagrams}(a)) is shown for
times $\Gamma t<1.5$, along with the Inchworm result (Fig. \ref{fig:method_diagrams}(c))
up to $\Gamma t=10$. Bottom panel: Error estimate of data in upper
panel showing an exponential increase of the error as a function of
time due to the dynamical sign problem in the bare method, and a roughly
constant error in the inchworm method.\label{fig:populations}}
\end{figure}

In the top panel of Fig.~\ref{fig:populations} we show the time-evolution
of the four populations (diagonal density matrix elements) after a
quench, as described by the bare hybridization expansion for times
$t\lesssim1.5$ (light lines) and by our inchworm algorithm (dark
lines). The system, initially in state $|\uparrow\rangle$, slowly
relaxes to a configuration in which $\uparrow$ and $\downarrow$
are degenerate. We observe that results for both numerically exact
algorithms agree within errors, but for $t\gtrsim1,$ bare QMC data
becomes noisy.

An error analysis (bottom panel) shows an exponential increase of
the bare error as a function of time (for the constant simulation
time per point used here), making times $\Gamma t\gg1$ inaccessible.
The large noise is a direct consequence of the dynamical sign problem.
In contrast, the error in the inchworm algorithm plateaus, allowing
access to significantly longer times. The inchworm error estimate
has been obtained from the standard deviations between completely
independent runs with uncorrelated statistical errors, thereby capturing
the full error propagation. The plateau of the noise implies that
the average sign stays constant as a function of time, and that there
is no observable error amplification due to repeated use of propagators
from earlier times.

\begin{figure}
\includegraphics[width=8.6cm]{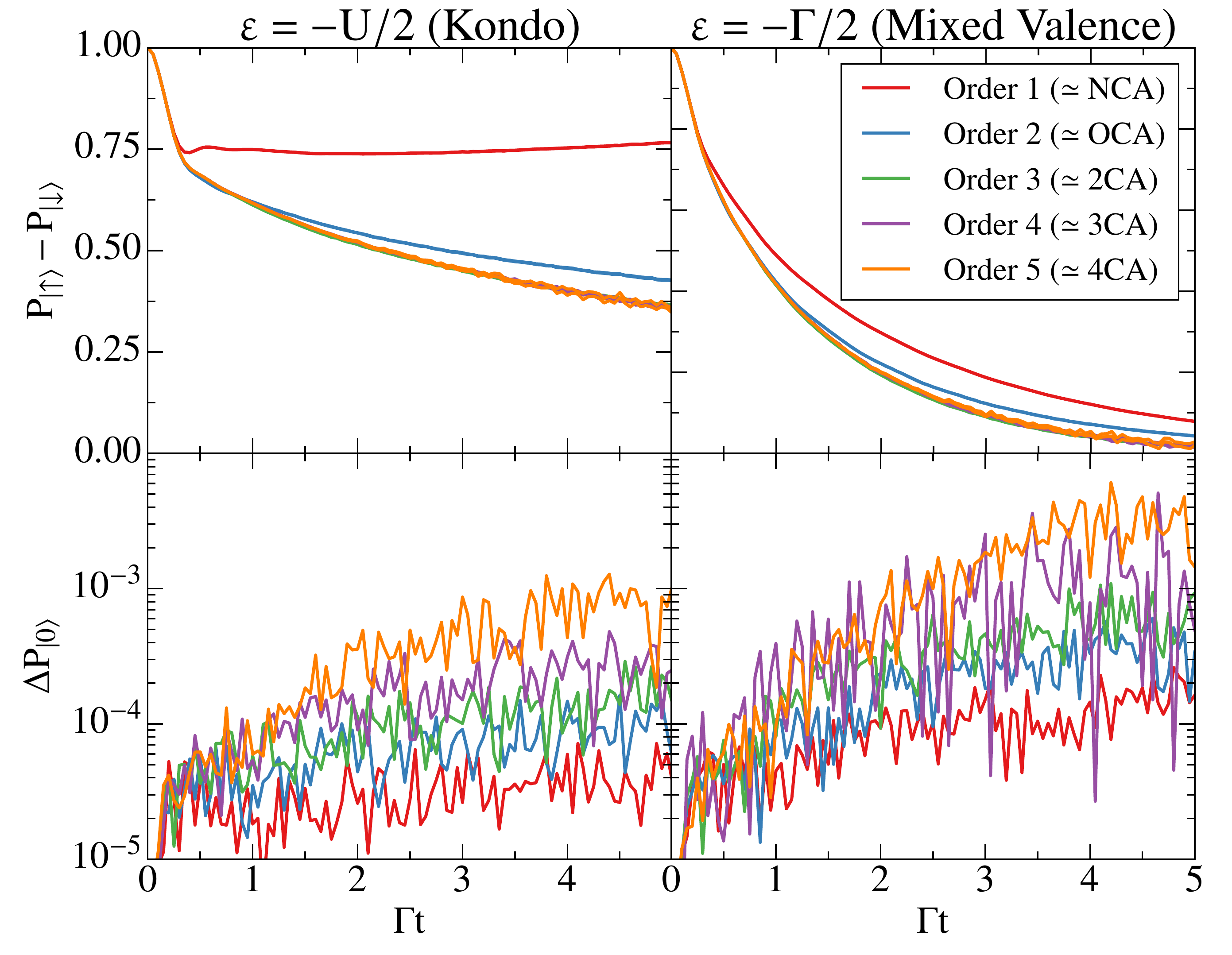}\caption{Top: population as a function of time after a coupling quench at $U=8\Gamma$
and$\beta\Gamma=50$, computed for a system in the Kondo regime (left
panels) and in the mixed valence regime (right panels). Different
traces show the convergence as a function of Inchworm expansion order.
Bottom: error estimate of the populations for different inchworm expansion
orders as a function of time.\label{fig:order_convergence}}
\end{figure}

To assess convergence with expansion order, we plot the magnetization
$P_{|\uparrow\rangle}-P_{|\downarrow\rangle}$as a function of time
in Fig.~\ref{fig:order_convergence}. The left panel shows parameters
in the Kondo regime $\varepsilon_{\sigma}=-U/2$, the right panel
parameters in the mixed valence regime $\varepsilon_{\sigma}=-\Gamma/2$.
Results of the inchworm method are exact only at infinite expansion
order. If the maximum expansion order is artificially restricted to
$1$, the relaxation to steady state is slow (right panel) or even
absent (left panel). As the maximum order is gradually increased,
the relaxation timescales shorten and (for these parameters) converge
at an expansion order of $\sim3-4$. In the limit $\Delta t\rightarrow0$
(we used a small but non-zero $\Delta t=0.05/\Gamma$), the diagrams
enumerated by the inchworm algorithm correspond to the NCA diagrams
for order 1, the OCA diagrams for order 2, the two-crossing diagrams
for order 3, etc. Fig.~\ref{fig:order_convergence} therefore shows
that at least a two-crossing approximation is required to correctly
capture the real-time evolution of this system.

The error analysis (bottom panels of Fig.~\ref{fig:order_convergence})
reveals that the error for each order first increases, then converges
to a constant, thereby overcoming the exponential scaling commonly
associated with a sign problem. The magnitude of the error increases
for increasing order, due to the larger sampling space available and
the fact that the calculations are performed at a fixed computational
cost, but because the error increases by an approximately constant
factor between any two orders, it may be eliminated by a small constant
increase in computer time (a factor of $\sim3$ in this case). This
graceful scaling, along with the rapid convergence to the exact result,
allows us to establish the algorithm as a numerically exact method.

While the same results could in principle be obtained by systematically
increasing the order of a semianalytical skeleton expansion (for example
improving the level of approximation from non-crossing to one-crossing
to two-crossing etc.), the computational expense typically increases
very rapidly level (for example each added crossing in an n-crossing
approximation adds a computational cost $\sim\left(\frac{t}{\Delta t}\right)^{2}$.
In practice, to our knowledge, non equilibrium calculations even at
the two-crossing level have been performed only to relatively short
time \cite{eckstein_nonequilibrium_2010}, and higher order calculations
have not been carried out. Fig.~\ref{fig:order_convergence} shows
that the inchworm algorithm can access the three- and four-crossing
approximations.

\begin{figure}
\includegraphics[width=8.6cm]{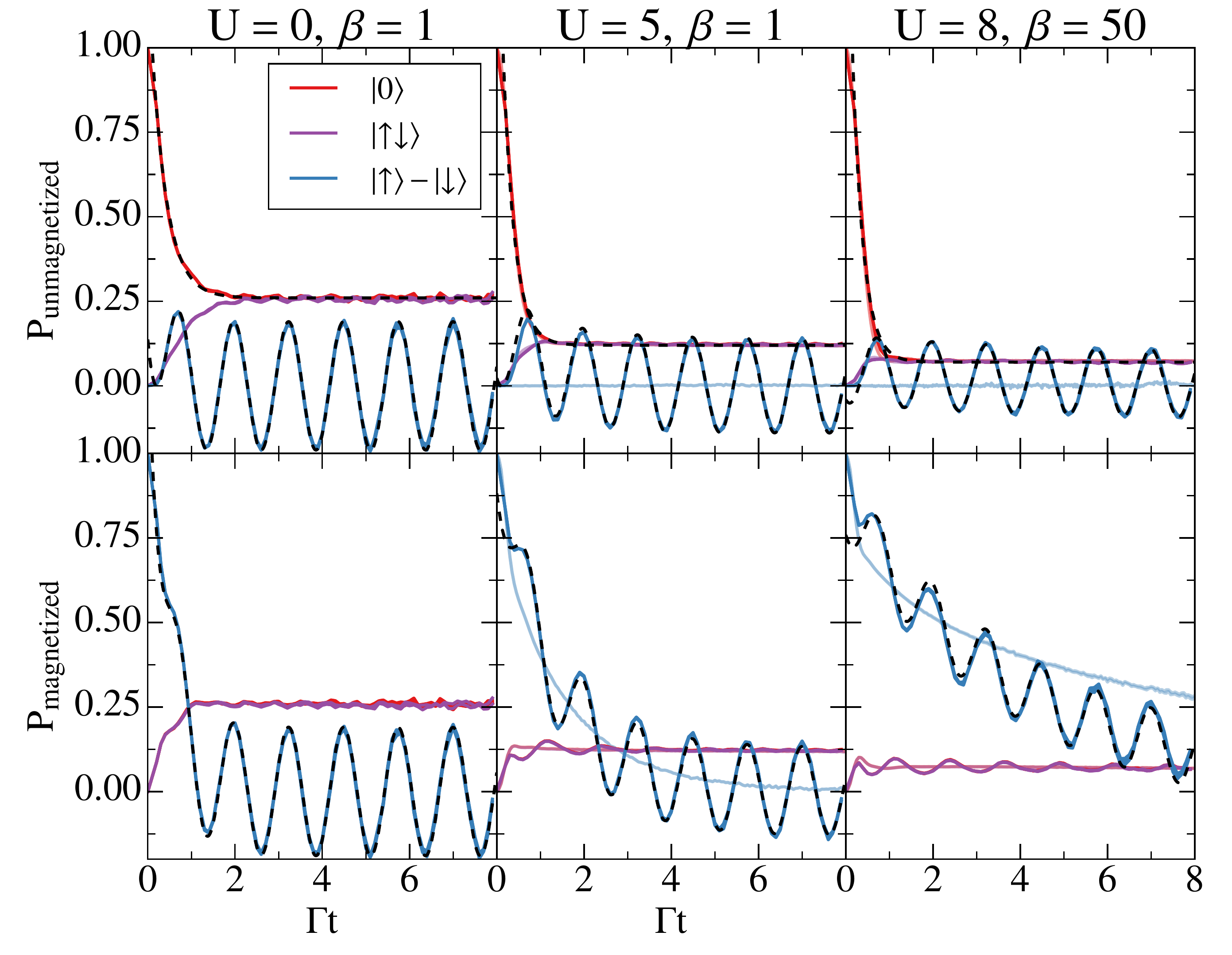}\caption{Population and magnetization dynamics of the quantum dot computed
at interaction strengths and temperatures shown in the presence of
a time dependent magnetic field $h(t)=2\Gamma\sin\left(\omega t\right)$
with $\omega=5\Gamma$. Top plots: dot initially in the empty state
$\left|0\right\rangle $. Bottom plots: dot initially in the fully
magnetized state $\left|\uparrow\right\rangle $. Underlying lighter
curves show the time evolution for $h=0$ with otherwise identical
parameters. Dashed black curves show fits to $f\left(t\right)=A+Be^{-\gamma t}+C\sin\left(\omega_{0}t+\phi\right)$.
In units where $\Gamma=1$, the charge relaxation rates ($\gamma$
for $\left|0\right\rangle $, $\left|\uparrow\downarrow\right\rangle $
) are $\gamma_{c}=$2.83, 3.8 and 4.0 for $(U,\beta)=(0,1),~(5.0,1)$
and $(5.8,50)$ respectively. The spin relaxation rates in the presence
of the field are $\gamma_{s}=$3.3, 0.81 and 0.25 (dot initially empty)
and 2.4, 0.81 and 0.25 (dot initially fully magnetized). The spin
relaxation rates for $h=0$ are 0.68 for $U=5$, $\beta=1$; and $0.11$
for $U=8,~\beta=50$. The final field amplitudes $C$ are 0.19, 0.13
and 0.1. In all cases, $\phi=-2.0$.\label{fig:magnetic_field}}
\end{figure}

In Fig.~\ref{fig:magnetic_field} we display the time dependence
of the probability that the dot is empty or doubly occupied (these
reflect the dot charge dynamics) and the magnetization, starting from
either an unmagnetized initial state (top panels) or a fully magnetized
initial state (bottom panels) and computed in the presence of an oscillating
magnetic field represented as a time and spin-dependent level shift
$\left(\varepsilon_{\uparrow}-\epsilon_{\downarrow}\right)\left(t\right)=2h\sin\left(\omega t\right)$.
Response to oscillating fields has been studied in the context of
currents induced by oscillating voltages\cite{nordlander_kondo_2000,zheng_kondo_2013}.
Current relaxation is rather fast even in the Kondo regime \cite{gull_numerically_2011},
so the numerical problems are less severe, but even in this case the
equation of motion methodology used in the more recent studies can
have convergence issues in the Kondo regime\cite{hartle_decoherence_2013,hartle_transport_2015}.
Here, we focus on the more challenging issue of the spin dynamics.
Three regimes are compared: the noninteracting case (left panel),
at the edge of the Kondo regime (center panel), and deeper in the
Kondo regime (right panel). As $U$ is increased and $T$ is decreased,
the charge relaxation time is shortened while the spin relaxation
time lengthens dramatically. We quantify the effects by fitting the
data to the simple phenomenological form $f\left(t\right)=A+Be^{-\gamma t}+C\sin\left(\omega_{0}t+\phi\right)$.
Fits are seen to be extremely good and reveal a more than factor of
10 increase in the spin lifetime and 50\% decrease in the charge lifetime
as the Kondo regime is entered, as well as an interesting dependence
of the spin relaxation time on the strength of the oscillating field.
A more detailed study of the spin dynamics, including an analysis
of the dependence on strength and frequency of the driving field,
will be presented elsewhere. 

In conclusion, we have presented a QMC method for real-time propagation
which we have termed the Inchworm algorithm, as it is based on gradually
`inching' along the Keldysh contour. The algorithm takes advantage
of previously computed propagation information by reusing it when
extending the propagation to longer times. We have implemented the
algorithm for the Anderson impurity model in the hybridization expansion,
where we were able to access slow spin dynamics in the strongly correlated
Kondo regime and observe its response to an oscillating magnetic field.
Our method suppresses the dynamical sign problem to such a degree
that the polynomially scaling part of the algorithm becomes dominant.
We also showed how high-order skeleton expansions are accessible by
truncating the expansion, at a scaling which is quadratic at any order
rather than being governed by a power law with the power proportional
to the order. 
\begin{acknowledgments}
The authors would like to thank Andrey Antipov and Yevgeny Bar Lev
for helpful comments and discussions. AJM and GC acknowledge support
from the Department of Energy under grant No. DE-SC0012375. EG acknowledges
support by DOE ER 46932. DR acknowledges support by NSF CHF 1464802. 
\end{acknowledgments}

 \bibliographystyle{apsrev4-1}
\bibliography{Library}

\end{document}